\documentclass[twocolumn,aps,preprint,letter,10pt]{revtex4}
\usepackage{graphicx,graphics,float}
\usepackage{dcolumn}
\usepackage{bm}
\usepackage{amsmath,amssymb}
\usepackage{mathrsfs}
\usepackage{xcolor}
\begin{document}
\title{The Fast Mode-Fourier-Transform and the bands
of the Bethe chain}
\author{Jose Reslen}\email{josereslen@uniatlantico.edu.co}
\affiliation{Coordinaci\'on de F\'{\i}sica, Universidad del
Atl\'antico, Carrera 30 N\'umero 8-49, Puerto Colombia.}%
%
%
\newcommand{\Keywords}[1]{\par\noindent{\small{\em Keywords\/}: #1}}
\begin{abstract} 
A many-body Fourier transformation with logarithmic scaling
in the number of necessary two-site gates is implemented.  The
protocol is applied to study the Bethe chain as a
prototype of a translationally invariant system. The resulting band
diagram features a flat pattern highly correlated to interaction
mechanisms. The introduced protocol can be applied to a wide
spectrum of scenarios and can offer new possibilities of simulation
and analysis.\\
\end{abstract}
\maketitle
\section{Introduction}
%
%
Science in general is driven by the development of methods and
techniques that make it possible to explore in the closeness of the
existing boundary of knowledge. In the numerical-method arena in
particular, a number of protocols have been used successfully over
the last decades in conjunction with the growing availability of
computational resources that characterize current technological
trends. In the realm of quantum physics, numerical protocols such
as those based on tensor renormalization \cite{mps1} or functional
optimization \cite{dft} have been applied extensively and with
various degrees of success and validity, allowing to extract vast
amounts of information and insight. One technique that has proven
versatile and far-reaching in many fields is the Fast Fourier
Transform (FFT) \cite{fft}. This technique has resulted specially
useful in cases where the representation must be efficiently
switched between two scalar domains, oftentimes Time and Frequency.
This generic conception by itself has been immensely prolific in
terms of applications and practical possibilities. This outlook
suggests an analogous mechanism is necessary when the transform
domains are made of operators instead of scalars. A particular case
where the relevance of this extension becomes apparent is in
many-body systems in which the switch between position and
momentum must be efficiently calculated, as for
instance, when the topological properties of a configuration of
interacting particles are intended to be studied through its band
structure. Moreover, the increasing development of information
technologies has uncovered the importance of protocols and
algorithms that could train models of artificial intelligence and
neural networks, in addition to the relevance that the efficient
calculation of functions has in proof-of-work schemes of
cryptography models. 

It is generally accepted that the numerical- and analytical-study
of quantum systems is hindered by the presence of interaction, so
that most methods are primarily conceived from the single-body
perspective.  The case of band theory is fairly representative of
this situation: The symmetry behind Bloch theorem is common to
single- and many-body systems, but the scheme is always formulated
in the one particle description.  Interaction dynamics has a
tendency to scramble Hilbert spaces in intricate ways, which does
not go well with the mechanisms behind existing formulations.  This
conundrum suggests the following question: Can these formulations
be adjusted to effectively solve many-body systems with single-body
symmetries? The concept aims at exploiting the regularities
stemming from the symmetry using the methods developed for
single-body systems, taking advantage of the fact that the symmetry
itself is determined by a single-body generator. This kind of
symmetry is quite common in models of interest, being Translation
Invariance perhaps the most significant in Solid State
configurations due to its role in the descriptive aspect of the
theory.  However, this proposal would not be practical without an
efficient application protocol, since the vector bases of many-body
systems display exponential growth. It is in this area where we
believe a fast transform proposal can make a contribution.

The proposal is tested in an interacting chain of fermions
introduced by Bethe in connection to his renown ansatz
\cite{bethe}. In comparison, the introduced method offers a simple
way to get the system's spectrum and depict it in a band diagram,
allowing to engage with a more panoramic view of the energy
distribution and get more detailed and concrete insight. It seems
reasonable that the same development could be achieved in other
configurations of interest.
\section{The Mode Fourier Transform}
\label{s09181}
Consider a Fourier-like transformation where the sample set is
replaced by a set of $N$ fermion ladder operators, in the way that
follows
\begin{gather}
\hat{f}_{j} = \frac{1}{\sqrt{N}} \sum_{k=1}^{N} W^{(j-1) (k-1)}
\hat{c}_{k}, \hspace{0.25 cm} j=1,2,\dots N,
\label{e07231}
\end{gather}
being $W=e^{\frac{2\pi i}{N}}$. Since the transformation is
unitary, anticommutation relations prevail: $\{ \hat{c}_j,
\hat{c}_k \}=\{ \hat{f}_j, \hat{f}_k \}=0$ and  $\{ \hat{c}_j,
\hat{c}_k^\dagger \}=\{ \hat{f}_j, \hat{f}_k^\dagger
\}=\delta_j^k$. The transformation can also be seen as a basis
change.  Even though standard fermions have been employed in this
formulation, the analysis that follows can be similarly applied to
bosons or Majorana fermions \cite{reslen5}.  The issue is that
although the transformation has the aspect of a Discrete Fourier
Transform (DFT), its functionality is different because the sample
set is not made of complex numbers but quantum operators.  They
cannot be added as scalars nor be altered independently.  However,
it is remarkable that they form a vector space under single-body
operations. One way how this property can be used to carry out the
Mode Fourier Transform (MFT) in equation (\ref{e07231}) is to
follow the reductionist approach implemented in reference
\cite{reslen7}.  For this, let us first focus on $\hat{f}_1$. In
this case all the factors of the transform are real but if they
were complex one could cancel the phases of all the coefficients by
means of local unitary transformations thus:
\begin{gather}
e^{-i\varphi \hat{c}_j^\dagger \hat{c}_j} \hat{c}_j e^{i\varphi
\hat{c}_j^\dagger \hat{c}_j} = e^{i\varphi} \hat{c}_j.
\label{e07252}
\end{gather}
Second, regard the following next-neighbor unitary transformation
\begin{gather}
\hat{U}_{j} = e^{-i\theta_{j} \hat{h}_j }, \hspace{0.25 cm}
\hat{h}_j = \frac{1}{2 i} (\hat{c}_{j+1}^\dagger \hat{c}_j -
\hat{c}_j^\dagger \hat{c}_{j+1}).
\label{e07251}
\end{gather}
The effect of this operation on the last two terms of the expansion
takes the following form
\begin{gather} 
\hat{U}_{N-1} \hat{f}_1 \hat{U}_{N-1}^{-1} = \hat{c}_1 + \hat{c}_2 + \dots \nonumber \\
+ \left ( \sin \left(
\frac{\theta_{N-1}}{2} \right) + \cos \left( \frac{\theta_{N-1}}{2}
\right)  \right )\hat{c}_{N-1} \nonumber \\
+ \left ( \cos \left(
\frac{\theta_{N-1}}{2} \right) - \sin \left( \frac{\theta_{N-1}}{2}
\right)  \right )\hat{c}_{N}.  
\end{gather}
In accordance, $\hat{c}_N$ can be canceled by setting
$\theta_{N-1}=\frac{\pi}{2}$. Notice that both local and two-site
transformations affect all the modes, not only $\hat{f}_1$. This is
a key difference with the scalar DFT, where each expansion can be
handled independently. The above cancellation can be repeated in
the same fashion using a transformation $\hat{U}_{N-2}$ and setting
an angle $\theta_{N-2}$ that precisely extinguish $\hat{c}_{N-1}$.
The process, often referred to as folding here, continues until
only $\hat{c}_1$ is left. At this stage, the factors of all the
other modes have changed, but there is one crucial feature: None of
the expansions apart from the folded $\hat{f}_1'$ contains
$\hat{c}_1$.  This must be so because the transformed operators
ought to comply with the same anticommutation relations than the
originals.  Consequently, the process can be repeated taking
$\hat{f}_2'$ with the new coefficients resulting from the folding
of $\hat{f}_1$ and applying two-site unitary transformations that
cancel elements sequentially until $\hat{c}_2$ alone is left. The
folding goes on until all the modes have been folded. As all the
transformations involved are unitary, it is possible to recover
expansion (\ref{e07231}) starting with the set of $\hat{c}_j$s and
applying the inverse transformations in reverse order. Since,
contrary to single-site operations, two-site operations can
increase the system's entropy, the protocol efficiency can be
measured by the number of two-site operations needed in total to
get the $\hat{f}_j$s from the $\hat{c}_j$s in the way just
described. The first folding takes $N-1$ gates, and for every new
reduction, one less gate is needed. Hence, there are
$\sum_{n=1}^{N-1} N-n = N(N-1)/2$ two-site transformations in
total. In the same way that the normal DFT has a more efficient
implementation, the Fast Fourier Transform (FFT) \cite{fft}, here
we explore the possibility of improving the MFT efficiency by way
of reducing the total number of two-site transformations.
\section{The Fast Mode-Fourier-Transform}
A landmark feature of the FFT is that in every step of
the protocol each pair of elements of the current series is used to
generate a corresponding pair of elements for the next series,
being the first series the original data and the final series their
DFT. Suppose two modes of the current series, $\hat{c}_a$ and
$\hat{c}_b$, are used to generate a pair of modes of the next
series, $\hat{f}_a$ and $\hat{f}_b$, through the following
identities:
\begin{gather}
\hat{f}_a = \hat{c}_a + W_n^k \hat{c}_b, \hspace{0.25 cm} \hat{f}_b = \hat{c}_a - W_n^k \hat{c}_b.
\end{gather}
The factor is given by $W_n=e^{\frac{2\pi i}{n}}$, being $n=2^l$ a
power of two that halve its value on each step of the decimation
process. It begins being $n=N$ and ends up being $n=2$. Integer $k$
determines which pair of the current set is being used to get a
new pair of the next set.  It goes from zero until half the number of
elements in the set minus one.  Indexes $a$ and $b$ do not need be
neighbors. To get the new modes one can use the same folding method
described in the previous section. It begins by wiping the complex
phase from factor $W_n^k$ using transformation (\ref{e07252}) on
mode $\hat{c}_b$, and then applying a two-site unitary
transformation analogous to (\ref{e07251}) to fold both modes
simultaneously. Then, $\hat{f}_a$ and $\hat{f}_b$ can be obtained
from $\hat{c}_a$ and $\hat{c}_b$ by reversely executing the folding
protocol using the inverse transformations. This pairing operation
is then implemented in the following way: The original set is
divided in two continuous halves, instead of in even and odd
distributions. Then the first element of the first half is paired
with the first element of the second half using $k=0$. Then the
second element of the first half is paired with the second element
of the second half using $k=1$, and so on. This constitutes the
first round of operations and leaves a new set of elements. This
new set is then divided in two halves in the same way as before,
and on each of them the pairing algorithm is repeated, each time
with half the total number of elements. The protocol continues as
long as more than one element remains after the last subdivision.
Finally, the mode indexes must be bit-reverse ordered to get the
correct distribution of factors.  As can be seen, there are $\log_2
N$ subdivisions, and for every subdivision $N/2$ two-site
operations must be applied. On total, there are $(N \log_2 N)/2$
two-site transformations involved, clearly a reduction in
comparison to the MFT. The whole transformation can be formulated
in the following terms
\begin{gather}
\hat{F} = \hat{B}_{RO} \prod_{l=\log_2 N}^1 \prod_{j=1}^{N/2^l} \prod_{k=0}^{(2^{l-1} - 1)}
e^{ i \pi(1 - 2^{-(l-1)}   k ) \hat{c}_{y}^\dagger \hat{c}_{y}}
\hat{U}_{x,y},
\label{e07261}
\end{gather}
where $x = 1 + k + (j-1) 2^{l}$ and $y = x+2^{l-1} = 1 + k + (2j-1)
2^{l-1}$. The two-site gate is given by
\begin{gather}
\hat{U}_{x,y} = e^{-\frac{\pi}{4}  (\hat{c}_x^\dagger \hat{c}_{y} -
\hat{c}_{y}^\dagger \hat{c}_x )}. 
\label{e07252}
\end{gather}
Since this transformation involves sites at different distances, it
is affected by sign assignments related to operator ordering.
Operator $\hat{B}_{RO}$ represents the bit-reverse ordering of mode
indexes. It is effectively realized as a swap of basis coefficients
plus a corresponding sign correction.  Both direct- and
inverse-FMFT can be implemented using virtually the same algorithm,
the only change being $W_n \rightarrow W_n^{-1}$.  

Since the FMFT requires two-site operations involving distant
neighbors, its implementation in terms of Matrix Product States
(MPS) is impractical. However, an efficient algorithm can be coded
using the standard Fock basis and exploiting the fact that, since
in every step of the transform only two sites of the chain are
involved, just a fraction of the states composing the whole basis
is needed at any given time. Another aspect of the procedure is
that, contrary to the protocol described in the previous section,
pairing operations are independent from each other. This offers the
chance to parallelize the algorithm. In addition, the procedure is
completely mechanical and systematic. These features constitute a
clear advantage with respect to the more general implementation
discussed before.
\section{Application to the Bethe chain}
\begin{figure}
\begin{center}
\includegraphics[width=0.45\textwidth,angle=0]{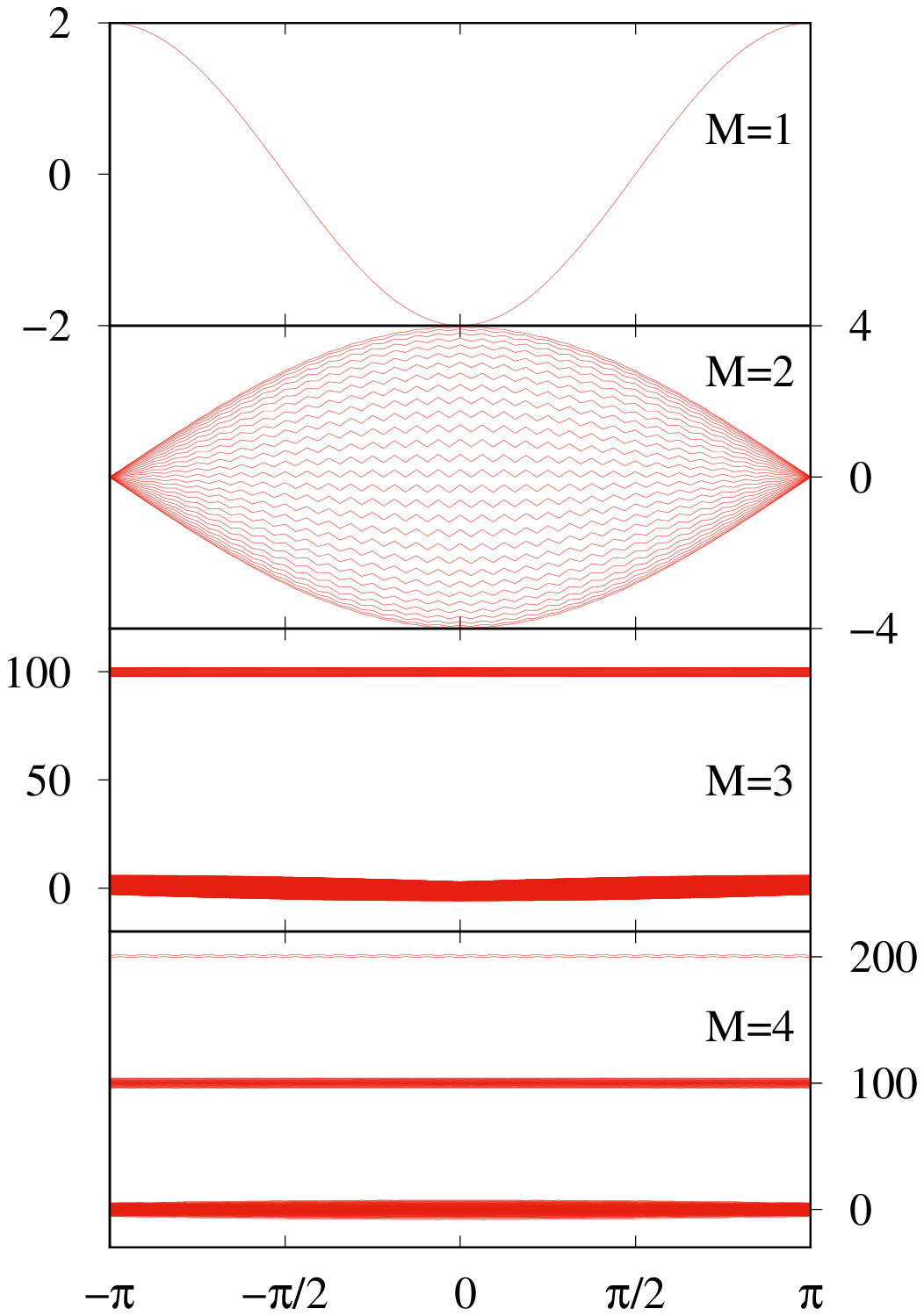}
\caption{Energy vs K in equation (\ref{e10131}) for N=64 and U=100.
Flat bands arise in relation to the number of particles. The
distance between bands is close to the interaction intensity, which
fits the view that each band is determined by states with a
corresponding number of particles in next-neighbors positions.}
\label{fig1}
\end{center}
\end{figure}
Let us consider a one-dimensional fermion chain exhibiting
next-neighbor hopping and interaction in the way established by the
following Hamiltonian
\begin{gather}
\hat{H} = \overbrace{-J \sum_{j=1}^N (\hat{c}_j^\dagger \hat{c}_{j+1} +
\hat{c}_{j+1}^\dagger \hat{c}_j)}^{\hat{H}_H} + \overbrace{U \sum_{j=1}^N
\hat{c}_{j+1}^\dagger \hat{c}_{j+1} \hat{c}_{j}^\dagger}^{\hat{H}_I} \hat{c}_{j}.
\label{e09181}
\end{gather}
Ladder modes operate in the space of Wannier functions describing
spinless fermions. These modes obey the same algebraic properties
than the modes introduced in section \ref{s09181}.  Integer $N$ is
the total number of sites and $M$, though not present in the
Hamiltonian, is the total number of fermions and is also a
preserved quantity. The chain displays periodic boundary conditions
$\hat{c}_{j+N} = \hat{c}_{j}$.  The system is invariant under
many-body time-reversal shifts and as such displays Krammers
degeneracy.  Constants $J$ and $U$ determine the intensity of
hopping and interaction respectively. Hereafter we make $J=1$ in
order to fix the energy unit and set the constants dimensionless.
The minus sign in the hopping term guarantees that states of low
energy correlate to low mobility in a way that is consistent with
the notion that such a term models kinetic energy. One way to
realize the scheme is to consider electrons in a ring, such as in a
Superconducting Quantum Interference Device.  This system is
effectively equivalent to the chain introduced by Bethe and as such
can be treated using the Bethe ansatz. A recent analysis of the
eigenstate structure has shown that the many-body nature of the
interaction incorporates long range correlations that cannot be
modeled as a product of neighbor gates, which makes the application
of MPS methods less suitable for this kind of systems in the
many-body regime \cite{reslen7}. The Hamiltonian commutes with the
Translation operator, $\hat{T}$, which {\it simultaneously} shifts
all modes in the same direction, like so
\begin{gather}
\hat{T} \hat{c}_j \hat{T}^{-1} = \hat{c}_{j+1}, \hspace{0.25 cm} \forall j.
\label{e09182}
\end{gather}
According to fundamental results from quantum theory, there exists
a common basis for $\hat{H}$ and $\hat{T}$. An eigenbasis of
$\hat{T}$ can be built using a product of $M$ modes, thus
\begin{gather}
|j_1,j_2,...,j_M \rangle =  \hat{f}_{j_1}^\dagger
\hat{f}_{j_2}^\dagger... \hat{f}_{j_M}^\dagger |0\rangle = \prod_{j} \hat{f}_{j}^\dagger |0\rangle,
\label{e09183}
\end{gather}
where $\hat{f}_{j}$ is given by equation (\ref{e07231}) and
$|0\rangle$ is the state with no fermions. Following the image of
electrons in a ring, the transformation amounts to shifting to an
angular momentum basis. Integers $j_1,j_2,...,j_M$ must be in
increasing order, but need not be consecutive. Using equation
(\ref{e09182}) it can be shown this eigenstate obeys the following
identity
\begin{gather}
|\hat{T}. j_1,j_2,...,j_M \rangle = e^{\frac{2 \pi i}{N} \sum_j
(j-1)}  |j_1,j_2,...,j_M \rangle \nonumber \\
= e^{i k}  |j_1,j_2,...,j_M \rangle.
\label{e09184}
\end{gather}
As a consequence, all eigenstates with values of $k$ for which 
\begin{gather}
K = mod(k-\pi,2\pi)
\label{e10131}
\end{gather}
deliver the same result will display the same eigenvalue. These
states form an independent subspace where the Hamiltonian can be
realized as a matrix. The spectrum of (\ref{e09181}) is thus
obtained as the collection of spectra from every subspace. The
value of $K$ can be used to characterize the corresponding subspace
in a way equivalent to the quasimomentum in band theory. As $K$ can
take $N$ different values and the Hilbert space dimension is
$N!/M!(N-M)!$, the dimension of a single subspace would be
approximately $(N-1)!/M!(N-M)!$. Similarly, any two-site operation
involves $(N-2)!/(M-1)!(N-M-1)!$ basis states.  Since the hopping
term is diagonal in the transformed basis, the problem is to find
the matrix elements of the interaction part.  This can be
accomplished utilizing operator (\ref{e07261}) in the next fashion
\begin{gather}
\langle j_1,j_2,...,j_M | \hat{H}_I. j_1,j_2,...,j_M \rangle =
\nonumber  \\
\langle \hat{F}. n_1,n_2,...,n_M | \hat{H}_I \hat{F}.
n_1,n_2,...,n_M \rangle = \nonumber  \\ \langle n_1,n_2,...,n_M |
\hat{F}^{-1} \hat{H}_I \hat{F}. n_1,n_2,...,n_M \rangle.
\label{e09191}
\end{gather}
Ket $| n_1,n_2,...,n_M \rangle $ is a Fock state in the space of
chain sites in which Hamiltonian (\ref{e09181}) has been
formulated. The protocol then consists in applying the FMFT on
every Fock state on each subspace, then the interaction operator, and
finally the inverse FMFT. The coefficients of the resulting
superposition correspond to the matrix elements of the interaction
operator. The hopping terms can be added as diagonal contributions.
The spectrum is then calculated by direct diagnolization over
independent subspaces.  

Figure \ref{fig1} shows the energy dependence as a function of $K$
for different occupation densities. Notice that the shown band
diagrams would not result naturally from a Bethe ansatz approach.
The case $M=1$ is consistent with the free particle scenario in
which the system would be a conductor as long as the Fermi energy
cross the band.  The case $M=2$ shows the influence of interaction
in the form of level splitting, but the effect is still moderate in
the sense that the conduction profile remains as in the previous
case. The cases $M=3$ and $M=4$ depict a more complex situation
where notoriously flat-bands emerge. The conduction properties will
swing between conductor and insulator as the Fermi energy is swept
from the bottom band upwards. The emergence of bands can be
explained by noticing that as fermions occupy periodic states they
recreate a periodic potential, akin of an effective lattice.  This
is experienced by other fermions in the system. The intensity of
the interaction will determine the height of this effective lattice
and will affect the range of energies that can be accessed by
particles.  This behavior can be used to engineer highly correlated
states dominated by the interaction and explore their topological
properties. The relation between the band pattern and system
parameters such as interaction or number of particles can be
exploited to calibrate the gap or the number of bands. In
particular, the band patterns shown in figure \ref{fig1} fit the
profile needed for a system to display Rabi oscillations or
Electromagnetically Induced Transparency for the cases $M=3$ and
$M=4$ respectively. One difference with the conventional
interpretation is that every point of the band represents a
many-body state. It is also noticeable in these same cases that the
distance between bands is approximately equal to the intensity of
the interaction.  This feature can be used to measure constant $U$
through a frequency-absorption characterization.  It also suggests
that every band displays a determined number of particles in
neighbor positions, being this an integer that indexes the band.
This feature could be used to design alternative wavefunction
ansatzs for the model. The bottom-band case is special in that its
interaction average is close to zero in spite of being strongly
influenced by interaction mechanisms through the Hamiltonian
couplings. In a low temperature scenario, this band fills first, so
that fermions will undergo a phase with little contribution to
dissipation via particle collisions.  Bandwidth depends on band
index too, becoming smaller the higher the band. Flat
bands with small bandwidth have potential applications as
conduction channels with negligible dissipation via lattice
collisions \cite{fb1,fb2}. It would be interesting to observe how
bands arise as a function of interaction and its potential relation
to a phase transition. 
\section{Conclusions}
An implementation of a Fourier Transform between field domains that
scales in logarithmic proportions have been introduced and used to
study the band pattern of a many-body system. The method displays
specific advantages with respect to a general formulation. The
application to the Bethe chain revealed the effect of interaction
in the band diagram, which manifests in the form of seemingly flat
bands that correlate to the number of occupied
next-neighbors-places in the band.  The system's spectrum can be
used to find thermodynamic averages and dissipation tendencies. It
seems possible that the introduced algorithm could be adjusted in
the direction of renormalization- or decimation-formulations in
order to make it more suitable to large-scale applications. A
natural extension of the current study is the application to the
Hubbard model. The effect of the spin and the particular
characteristics of that model might give rise to topological
features.  Similarly, it would be interesting to explore how the
FMFT could be used to study non-unitary dynamics in open quantum
systems \cite{reslen6,reslen8}.
\end{document}